\documentclass[conference]{IEEEtran}
\IEEEoverridecommandlockouts
\usepackage{cite}
\usepackage{amsmath,amssymb,amsfonts}
\usepackage{algorithmic}
\usepackage{graphicx}
\usepackage{float}
\usepackage{textcomp}
\usepackage[hidelinks=true]{hyperref}
\usepackage{xcolor}

\def\BibTeX{{\rm B\kern-.05em{\sc i\kern-.025em b}\kern-.08em
    T\kern-.1667em\lower.7ex\hbox{E}\kern-.125emX}}
\begin{document}

\title{Augmenting Medical Imaging: A Comprehensive Catalogue of 65 Techniques for Enhanced Data Analysis\\}


\author{\IEEEauthorblockN{ Manuel Cossio MMed, MEng}
\IEEEauthorblockA{\textit{Dept. of Mathematics and Computer Science} \\
\textit{Universitat de Barcelona}\\
Barcelona, Spain \\
manuel.cossio@ub.edu}

}

\maketitle

\begin{abstract}

In the realm of medical imaging, the training of machine learning models necessitates a large and varied training dataset to ensure robustness and interoperability. However, acquiring such diverse and heterogeneous data can be difficult due to the need for expert labeling of each image and privacy concerns associated with medical data. To circumvent these challenges, data augmentation has emerged as a promising and cost-effective technique for increasing the size and diversity of the training dataset. In this study, we provide a comprehensive review of the specific data augmentation techniques employed in medical imaging and explore their benefits. We conducted an in-depth study of all data augmentation techniques used in medical imaging, identifying 11 different purposes and collecting 65 distinct techniques. The techniques were operationalized into spatial transformation-based, color and contrast adjustment-based, noise-based, deformation-based, data mixing-based, filters and mask-based, division-based, multi-scale and multi-view-based, and meta-learning-based categories. We observed that some techniques require manual specification of all parameters, while others rely on automation to adjust the type and magnitude of augmentation based on task requirements. The utilization of these techniques enables the development of more robust models that can be applied in domains with limited or challenging data availability. It is expected that the list of available techniques will expand in the future, providing researchers with additional options to consider.

\end{abstract}


\tableofcontents

\section{Introduction}

The robustness and interoperability of machine learning models is heavily reliant on the size and variability of the training data. Incorporating diverse and heterogeneous data in the training process can reduce the risk of overfitting and enhance the model's ability to generalize to out-of-sample data. However, this is particularly challenging in medical imaging due to the need for expert labeling of each image, as well as the sensitive nature of the data and associated privacy concerns. As a result, the process of obtaining necessary permissions and implementing appropriate protection measures can be time-consuming and expensive. Data augmentation has emerged as a cost-effective, efficient, and accurate means of increasing the size and diversity of the training dataset. In this study, we explore the benefits of data augmentation in medical imaging and provide a comprehensive review of the specific techniques employed in this domain.

\section{Purposes of data augmentation in medical imaging}

\subsection{Increase the size of the dataset}
Data augmentation techniques allow for the creation of new samples that are similar to the original ones, which can increase the size of the training dataset. This is especially useful when the original dataset is small or imbalanced, which is often the case in medical imaging datasets\cite{mikolajczyk2018data}.

\subsection{Improve model generalization}
Data augmentation techniques can help to reduce overfitting by training the model on a more diverse set of examples. By exposing the model to a wider range of variations in the data, it can learn to generalize better and perform better on new, unseen examples\cite{zhao2020maximum}.

\subsection{Improve model robustness}
 By applying different types of data augmentation, the model can learn to recognize and extract important features regardless of changes in the input image, such as different rotations, translations, or lighting conditions. This can make the model more robust to variations in the data and more reliable in real-world scenarios\cite{rebuffi2021data}.

\subsection{Improve model performance}
 In some cases, applying data augmentation can directly improve the performance of the model on the task at hand. For example, applying color normalization or histogram equalization can improve the contrast and clarity of images, making it easier for the model to distinguish between different features\cite{poojary2021effect}.

\subsection{Improve model accuracy}
In some cases, applying data augmentation can directly improve the accuracy of the model on the task at hand. For example, applying rotation and flipping can help the model to recognize features that are oriented in different directions, which can be important for certain medical imaging tasks\cite{poojary2021effect}.

\subsection{Mitigate class imbalance}
Data augmentation can be used to balance the distribution of different classes in the dataset. For example, in medical imaging datasets, certain classes may be underrepresented, such as rare diseases or subtypes of cancer. By augmenting the examples in these classes, the model can be trained on a more balanced dataset, which can improve its performance on these classes\cite{afzal2019data}.

\subsection{Reduce the need for manual labeling}
Data augmentation can be used to simulate variations in the data that are difficult or expensive to collect manually. For example, in some cases it may be difficult to obtain images of a particular disease in different stages or from different imaging modalities. By applying data augmentation to the existing data, it is possible to simulate these variations and train the model on a more diverse set of examples without the need for additional manual labeling\cite{shorten2019survey}.

\subsection{Reduce domain shift in transfer learning}
Data augmentation can also be leveraged to standardize the training conditions in transfer learning. During pre-training on a general dataset, a specific set of data augmentation techniques can be applied to generate additional training examples. By utilizing the same set of techniques during the fine-tuning of the model with the pathology dataset, the model can be adapted more effectively to the new dataset. In this context, data augmentation can be considered a valuable tool for reducing the domain shift between the general dataset and the pathology dataset, thereby enhancing the model's ability to generalize to new and unseen data\cite{brion2021domain}.

\subsection{Increase robustness to artifacts and noise}
Medical images are often subject to artifacts and noise, which can degrade the performance of deep learning models. By applying data augmentation techniques that introduce similar artifacts and noise to the training data, the model can become more robust to these effects and better able to generalize to new, unseen images\cite{hussain2017differential}.

\subsection{Reduce overfitting}
Overfitting occurs when a model is trained too well on the training dataset, to the point where it memorizes the examples and fails to generalize to new, unseen data. Data augmentation can be used to introduce variability to the training data, which can help to prevent overfitting and improve the model's generalization ability\cite{perez2017effectiveness}.

\subsection{Enhance visualization and interpretation}
Data augmentation can be used to generate new images that can be used for visualization and interpretation purposes. For example, by applying color transformations to medical images, it is possible to highlight different regions or features that may not be visible in the original image\cite{tien2021cone}.

\subsection{Generate synthetic data}
Data augmentation can be used to generate entirely synthetic data, which can be useful in situations where the original dataset is limited or of poor quality. For example, generative models such as Generative Adversarial Networks (GANs) can be used to generate synthetic medical images that are visually realistic and can be used to augment the original dataset\cite{dimitriadis2022enhancing}.

\section{Data augmentation applied to medical imaging}

As highlighted in the previous section, data augmentation techniques have been shown to have a positive impact on various aspects of model training in medical imaging. Accordingly, in the subsequent lines, we aim to present a comprehensive catalog of data augmentation techniques that have been applied in this domain.

\section{Spatial transformation based}

\subsection{Random cropping}

Random cropping is a popular data augmentation technique that involves randomly selecting a portion of an image and using it as a new image. During the training process, the model is exposed to different portions of the original image, allowing it to learn features and patterns that are present across the entire image. To apply random cropping, a fixed-size window is randomly positioned over the original image, and the pixels within the window are used as a new image. This process can be repeated multiple times for each image to create additional training examples. The size of the window and the number of times the process is repeated can be adjusted based on the characteristics of the dataset and the desired level of data augmentation\cite{takahashi2019data, hao2021comprehensive}.

\subsection{Rotation}

Rotation is a common data augmentation technique used in computational pathology, where images are rotated by a certain angle. By rotating the image, the model becomes more robust to variations in orientation, which can be particularly useful when dealing with data that has varying degrees of rotation, such as tissue samples or microscope slides. To apply rotation, the original image is rotated by a random angle, typically within a predefined range. The angle of rotation can be selected based on the characteristics of the dataset and the desired level of data augmentation. For instance, if our dataset includes images of tissue samples that are originally rotated in different orientations, a larger range of rotation angles may be used to increase the diversity of the training data\cite{hussain2017differential, xu2020automatic, loey2020deep, deoliveira2021classification}.

\subsubsection{Random rotation}

A variation of the above technique is random rotation, where the angle of rotation is chosen at random. To do this, the minimum and maximum values must first be set and at each iteration a different value will be chosen at random\cite{sirazitdinov2019data, feng2020deep}.

\subsection{Flipping}

This technique, also known as mirroring, involves flipping the image horizontally or vertically, which helps the model become more robust to variations in image direction. Horizontal flipping involves flipping the image from left to right, while vertical flipping involves flipping the image from top to bottom. Both types of flipping can be used in combination to further increase the diversity of the training data\cite{hussain2017differential, feng2020deep, monshi2021covidxraynet, loey2020deep}.

\subsection{Scaling}

Scaling is a data augmentation technique that involves changing the size of an image, either by increasing or decreasing its dimensions. There are several ways to apply scaling to images, depending on the characteristics of the dataset and the desired level of data augmentation. One common approach is to randomly scale the images within a certain range during training. For example, the image could be randomly scaled up or down by a factor of 10\%, which would result in a slightly larger or smaller image. Scaling can also be used to standardize the size of images in a dataset. For example, if a dataset contains images of different sizes, they can be resized to a fixed size using scaling. This ensures that all images are the same size and simplifies the training process for the model\cite{chlap2021review, xu2020automatic}.

\subsection{Translation}

Translation is a data augmentation technique that involves moving an image in different directions, either horizontally or vertically. This is done by shifting the pixels in the image by a certain amount of pixels in a particular direction. For instance, when the image is translated horizontally to the right, a new space (typically black) is introduced on the left side of the image. The extent of this space increases as the amount of translation increases. By moving the image in this manner, the model can learn to recognize features of the image regardless of their position within the frame, which enhances the model's ability to handle variations in image position. Using another example, if an image of a tissue sample is slightly off-center, the model trained on such data might perform poorly on other off-center images. By using translation, the model can be trained on images that have been shifted in various directions, making it more robust to variations in image position\cite{chlap2021review}.

\subsection{Shearing}

Shearing is a type of data augmentation technique that involves displacing one part of an image relative to the other, causing it to become skewed. This is done by shifting one row of pixels in the image horizontally while keeping the other rows fixed, or by shifting one column of pixels vertically while keeping the other columns fixed. For example, when tissue is cut and mounted onto a slide, it can become stretched or compressed in certain areas, resulting in irregular shapes and structures. Shearing can simulate these types of deformations by skewing the image in a controlled manner. This can help a model learn to recognize these irregularities and make more accurate predictions on similar images in the future\cite{hussain2017differential}.

\subsection{Zooming}\label{zooming}

Zooming is a common data augmentation technique in computational pathology. Zooming in or out of an image involves changing its size, either by enlarging or reducing its dimensions, while keeping the content of the image centered.

In the context of computational pathology, zooming can help the model become more robust to variations in image scale. This is important because histopathological images can come from different sources, such as different scanners or microscopes, and can have varying sizes. By applying zoom as a data augmentation technique, the model can learn to recognize patterns at different scales and become more accurate at detecting features in images of varying sizes.

Zooming can be performed in different ways, such as by cropping and resizing the image or by using a zoom function that rescales the image while preserving its aspect ratio. It is important to note that excessive zooming can lead to loss of information and can negatively impact model performance. Therefore, careful selection of the amount and type of zoom is crucial to ensure optimal model performance \cite{monshi2021covidxraynet, loey2020deep,zeiser2020segmentation}.

\subsection{Perspective}

Perspective transformation is a data augmentation technique that applies a geometric distortion to an image. It is used to simulate the effect of viewing an object from different perspectives. This technique is particularly useful for tasks such as object detection, where the algorithm needs to recognize objects that may appear in different orientations and angles. Perspective transformation involves selecting four points in the input image and four corresponding points in the output image. These points define a perspective transformation matrix, which is then used to warp the input image to the desired output shape. The perspective transformation matrix is calculated based on the locations of the four points and the desired output shape. During the transformation, the algorithm applies a skewing effect to the image, making it appear as if it is viewed from a different angle. This results in the creation of new data samples with a similar appearance to the original samples, but with different perspectives. This technique can help improve the robustness of machine learning models by allowing them to recognize objects from different viewpoints\cite{zhong2021deep, wang2019perspective, li2021multi}.

\section{Color and constrast adjustment based}

\subsection{Brightness}

Brightness adjustment is a data augmentation technique used to modify the overall brightness of the image by increasing or decreasing its intensity. This technique can help the model become more robust to variations in lighting conditions that can affect the image quality. Brightness adjustment can be achieved by applying a linear transformation to the intensity values of the image, where the brightness of the image is increased or decreased by multiplying the intensity values by a constant factor.

For instance, if we want to increase the brightness of the image, we can multiply the intensity values by a factor greater than one. On the other hand, if we want to decrease the brightness of the image, we can multiply the intensity values by a factor less than one. By adjusting the brightness of the image, we can generate new images that are visually similar to the original ones, but have different brightness levels, which can help the model learn to recognize objects under different lighting conditions\cite{sirazitdinov2019data, tellez2019quantifying}. 

Brightness variations are a prevalent issue in whole slide images (WSIs) that have been scanned with different scanners. To overcome this challenge, brightness adjustment can be employed to enhance the model's performance in analyzing slides from diverse hospitals that may have been scanned using different equipment. By adjusting the brightness of the image, the model can learn to recognize features under various lighting conditions, leading to improved accuracy in analyzing WSI datasets. Therefore, employing data augmentation techniques, such as brightness adjustment, can aid in reducing the impact of brightness variations and enable the model to perform effectively on diverse WSI datasets\cite{tellez2019quantifying}.

\subsection{Contrast }

Contrast adjustment is a data augmentation technique that alters the contrast of an image by adjusting the difference between the brightest and darkest pixels. This can be achieved through various methods such as histogram equalization, which redistributes the pixel intensity values to improve the contrast of an image. Contrast adjustment can help the model become more robust to variations in image quality, particularly when the contrast of images in the dataset varies significantly. By applying contrast adjustment, the model can learn to recognize features in images with different contrast levels and improve its performance on images with poor contrast. Additionally, this technique can also be used to reduce the effect of noise in low-contrast images and enhance the visual appearance of certain features, making them easier to detect and classify by the model\cite{sirazitdinov2019data, tellez2019quantifying, zhong2021deep}.

\subsection{Gamma correction}

Gamma correction is a data augmentation technique that involves modifying the intensity of an image by adjusting the gamma value. Gamma is a non-linear function that is used to encode and decode the luminance or brightness of an image. In gamma correction, the gamma value is adjusted to modify the overall brightness of the image. This technique is particularly useful when dealing with images that have low contrast or when the lighting conditions are not ideal. By adjusting the gamma value, it is possible to enhance the contrast of the image, making it easier for the model to distinguish between different structures and features. Gamma correction can be applied in a variety of ways, such as globally to the entire image or locally to specific regions of interest\cite{sirazitdinov2019data, xu2020automatic, sun2021robust, zhong2021deep}.

\subsection{Difference among brightness, contrast and gamma correction}

As we saw in the two previous sections, brightness and contrast changes, and gamma correction are three different image processing techniques, although they can be used to adjust the overall luminance of an image.

Brightness adjustment involves changing the intensity of all pixels in an image uniformly, either by adding or subtracting a constant value to the pixel values or by multiplying them by a scaling factor. This changes the overall brightness of the image, making it brighter or darker.

Contrast adjustment involves changing the range of pixel intensities in an image, which can affect the difference in brightness between the lightest and darkest parts of the image. This is achieved by applying a linear transformation to the pixel values, which maps the original intensity range to a new range. Increasing contrast makes the bright parts of the image brighter and the dark parts darker, while decreasing contrast has the opposite effect

Gamma correction, on the other hand, involves non-linearly transforming the pixel values to adjust the brightness and contrast of an image. It works by raising each pixel value to a certain power (the gamma value) to obtain a new pixel value. By adjusting the gamma value, the dark or bright areas of an image can be enhanced or suppressed. 

\subsection{Hue adjustment}\label{Hue_adj}

Hue adjustment is a data augmentation technique that involves modifying the colors in an image by shifting the hue component. The hue component refers to the dominant color in an image, such as red, blue, or green. This technique can be used to generate variations in color that the model may encounter when analyzing real-world images. For example, variations in tissue staining during slide preparation can result in different hues in the same type of tissue. During hue adjustment, the hue component is shifted by a certain amount, resulting in a different overall color for the image. This technique can be applied either uniformly across the entire image or selectively to specific regions of interest. In addition to introducing color variations, hue adjustment can also be used to correct color imbalances in an image, such as those caused by lighting conditions or camera settings\cite{kora2020evaluation, tellez2019quantifying}.

\subsection{Saturation}

Saturation adjustment is another data augmentation technique. It involves modifying the saturation of an image, which refers to the intensity or purity of its colors. A fully saturated image has pure colors, while a desaturated image appears more washed out or gray. 

Saturation adjustment can be achieved by multiplying the color channel values by a scalar factor. A scalar factor greater than 1 increases the saturation of an image, while a scalar factor between 0 and 1 decreases the saturation. This technique can be useful in situations where images may have different levels of saturation due to variations in lighting conditions or color capture methods\cite{chen2020enhancement, naglah2022conditional}.

\subsection{Color jitter}

Color jitter is a data augmentation technique that involves randomly perturbing the color of an image by making small changes to its hue, saturation, brightness, and contrast values. By introducing random variations in the color of an image, color jitter can help a model become more robust to variations in color that may be present in different images of the same object or tissue. For example, color variations may be caused by differences in lighting conditions or staining procedures used to prepare tissue samples. By applying color jitter to an image, the model can learn to better recognize and classify the object or tissue despite these variations in color\cite{hussain2017differential}.

\subsection{Sharpening}

Sharpening is a data augmentation technique used to enhance the edges and details in an image. It works by increasing the contrast of the pixels surrounding edges, making them appear more pronounced. This can help the model become more robust to variations in tissue structure, as it can enhance subtle details and edges that may be important for classification or segmentation tasks.

It is important to note that while sharpening can be a useful tool for enhancing images, it can also introduce artifacts or noise into the image if not applied carefully. Therefore, it is important to evaluate the effect of sharpening on the specific dataset and task at hand before using it as a data augmentation technique\cite{zhong2021deep}.

\subsection{Color space transformation}

Color space transformation is a data augmentation technique that involves converting an image from one color space to another. Color space refers to the representation of color values in an image, which can affect the way colors are perceived and the amount of information available for analysis. 

Some common color spaces used in image processing include RGB (Red, Green, Blue), HSV (Hue, Saturation, Value), and LAB (Lightness, A, B). Each color space has its own advantages and disadvantages in terms of color representation, and certain colors may be better represented in one color space compared to another.

\subsubsection{HSV}

HSV stands for Hue, Saturation, and Value, which are three components that make up the HSV color model. Hue represents the actual color of the pixel, ranging from 0 to 360 degrees around a color wheel. It's often described as a color's "shade" or "tint". Saturation represents the intensity of the color, ranging from 0 (completely unsaturated, grayscale) to 100 (fully saturated). When saturation is increased, colors become more vivid, while decreasing saturation results in more pastel-like colors. Value represents the brightness or lightness of the color, ranging from 0 (black) to 100 (white). Increasing the value results in a brighter and more washed-out color, while decreasing the value makes the color darker\cite{marini2023data}.

In image processing, HSV color transformation can be used to adjust the color of an image while maintaining its brightness and contrast. For example, by increasing the saturation of an image, colors can be made more vibrant and stand out more. By changing the hue, colors can be shifted to different parts of the color wheel, allowing for interesting artistic effects or to correct for color casts in images. Finally, changing the value can adjust the overall brightness of an image without changing the underlying color information.

\subsubsection{Hue or HSV?}

Hue adjustment augmentations (see Section \ref{Hue_adj}) and HSV color space augmentations are different concepts. In hue augmentations, only the hue component changes while the saturation and value components remain constant. On the other hand, in HSV augmentations, all three components can be modified to create new images in the color space.

\subsubsection{LAB}

LAB color transformation is also a color space commonly used in image processing and computer vision. Unlike RGB color space, which represents color based on the intensity of red, green, and blue values, LAB color space represents color based on a combination of lightness, a, and b values. The L component, or lightness, represents the perceived brightness of the color, ranging from 0 (black) to 100 (white). The a and b components represent the color channels, with a ranging from green (-128) to red (+128) and b ranging from blue (-128) to yellow (+128). The LAB color space is designed to be perceptually uniform, meaning that a small change in the LAB values corresponds to a small change in the perceived color. This makes it a useful color space for applications such as image segmentation and color-based object detection\cite{vesal2018classification, veta2015assessment}.

\subsection{Color inversion}

Color inversion is a simple but effective data augmentation technique in which the colors of the image are inverted by subtracting the value of each pixel from the maximum value that can be represented by the image. For example, if the image is represented by 8-bit color depth (256 levels of intensity), the maximum value is 255, so each pixel value is subtracted from 255. This technique can help the model become more robust to variations in color representation. In some cases, the colors of the same object or tissue structure can appear different due to lighting conditions or image capture settings. Inverting the colors of the image can help the model learn to recognize the object or tissue structure regardless of its color\cite{talukdar2018data}.

\subsection{Histogram specification}

Histogram specification, also known as histogram matching, is a technique used to adjust the pixel value distribution of an image based on a reference histogram. The reference histogram can be either a pre-defined histogram or the histogram of another image. The goal is to transform the image so that its histogram matches the reference histogram, thereby equalizing the distribution of pixel values in both images.

The process of histogram specification involves several steps. First, the histogram of the input image is calculated, which represents the frequency distribution of pixel values. Then, the cumulative distribution function (CDF) of the input image is computed. The CDF is a function that maps each pixel value to its cumulative frequency in the image.

Next, the CDF of the reference histogram is calculated. The CDF of the reference histogram is used to compute a mapping function that transforms the pixel values in the input image to new values that better match the reference histogram. This mapping function is then applied to each pixel in the image, resulting in an image whose histogram matches the reference histogram\cite{jois2021boosting, fan2021mammography, cao2021novel}.

\subsection{Global contrast normalization}

Global contrast normalization (GCN) is a data augmentation technique commonly used in deep learning applications for image classification tasks. The goal of GCN is to normalize the overall intensity of an image, so that each pixel value falls within a certain range, typically between 0 and 1. This normalization process helps to reduce the effects of lighting variations and other noise factors that may exist in the original dataset. The normalization process involves two steps: mean subtraction and division by standard deviation. In mean subtraction, the average pixel value of the image is calculated and subtracted from each pixel value in the image. This step centers the pixel values around zero. Then, the resulting image is divided by the standard deviation of the pixel values, which scales the values to have a similar distribution. This technique helps to improve the model's ability to generalize to new images with different lighting and contrast conditions. However, it is important to note that GCN may not always be suitable for all types of datasets and tasks, as it can also introduce unwanted artifacts and distortions in some cases\cite{chougrad2018deep, arevalo2016representation}.

\subsection{Local contrast normalization}

Local contrast normalization is a data augmentation technique that aims to increase the contrast and enhance the edges of an image by normalizing the contrast within a small region of the image. Unlike global contrast normalization, which normalizes the contrast of the entire image, local contrast normalization operates on a smaller scale by dividing the image into small patches or regions and normalizing the contrast within each patch independently. Compared to global contrast normalization, local contrast normalization is more effective at enhancing local image features and preserving the overall structure of the image. This is particularly useful in medical imaging applications, where small structures and subtle features can be important for accurate diagnosis. Additionally, local contrast normalization can be applied to images with non-uniform illumination, where global contrast normalization may not be effective. However, local contrast normalization can be computationally expensive, especially when applied to large images or high-resolution datasets, and may require careful tuning of the patch size and normalization parameters\cite{arevalo2016representation}.

\subsection{Histogram equalization}

Histogram equalization is a data augmentation technique used to adjust the contrast of an image by redistributing the pixel intensities. In an image, the histogram represents the distribution of pixel intensities, with the horizontal axis showing the intensity values and the vertical axis showing the number of pixels with that intensity. A histogram with a narrow peak indicates that the image has a low contrast, while a histogram with a wide spread indicates a high contrast. Histogram equalization works by first calculating the cumulative distribution function (CDF) of the histogram, which gives the probability of a pixel having an intensity value less than or equal to a given value. The CDF is then used to map the intensity values in the original image to new values that are spread more evenly across the entire intensity range. This results in an image with a more uniform distribution of intensities, and therefore a higher contrast\cite{akarsu2016fast, zhong2021deep}. Histogram equalization can be applied to grayscale as well as color images. However, in color images, it is important to equalize the histograms of each color channel separately to avoid introducing color artifacts. Histogram equalization can be used as a data augmentation technique to generate new images with different contrast levels that can help the model become more robust to variations in image quality.

\section{Noise based}

\subsection{Gaussian noise}
Gaussian noise is a type of noise that is added to the image by introducing random values drawn from a Gaussian distribution. The addition of Gaussian noise to an image can help the model become more robust to variations in image quality, as it simulates the noise that can occur during image acquisition and preprocessing. The amount of noise added to the image can be controlled by adjusting the standard deviation of the Gaussian distribution. A higher standard deviation will result in more noise being added to the image\cite{hussain2017differential, sirazitdinov2019data, tellez2019quantifying, zhong2021deep}.

\subsection{Speckle noise}

Speckle noise is a type of noise that appears in images acquired by coherent imaging systems, such as ultrasound. It is caused by the interference of the wavefronts reflected from the different scatterers in the imaged object. Speckle noise appears as a grainy pattern with a granular texture that reduces the contrast and obscures the details of the image. As a data augmentation technique, speckle noise can be added to the image by multiplying the image by a random value drawn from a speckle distribution. This results in a noisy version of the original image that can help the model become more robust to variations in image quality. Speckle noise can be particularly useful in training models for medical image analysis tasks, such as ultrasound or OCT (optical coherence tomography) imaging, where speckle noise is a common source of image degradation\cite{lee2021principled}.

\subsection{Salt and pepper noise}

Salt and pepper noise is a type of noise commonly seen in digital images. It is named after the appearance of small white and black dots that resemble grains of salt and pepper sprinkled on an image. The noise is caused by random variations in the image signal during the image acquisition or transmission process.

In image processing, salt and pepper noise can be added as a form of data augmentation. This involves randomly changing some pixels in the image to either the minimum (black) or maximum (white) pixel values. The amount of noise added can be controlled by specifying the probability of each pixel being affected\cite{li2020identification, deoliveira2021classification}.

\subsection{Poisson noise}

Poisson noise is a type of noise that is commonly seen in digital images captured using low-light conditions or low dose radiographic imaging, where the number of photons reaching the detector is limited. Poisson noise is modeled using the Poisson distribution, which describes the probability of a given number of events occurring in a fixed interval of time or space. In image processing, Poisson noise manifests as grainy or speckled patterns on the image, especially in areas with low signal intensity. Adding Poisson noise to an image can simulate such noise in real-world scenarios and can help the model become more robust to variations in image quality. In the context of data augmentation, Poisson noise can be added to an image by modeling the noise using the Poisson distribution and adding a random value drawn from this distribution to each pixel value of the image. The amount of Poisson noise added to an image is usually controlled using a parameter called the noise level, which determines the variance of the Poisson distribution\cite{zhovannik2019learning, kohlberger2019whole}.

\section{Deformation based}

\subsection{Elastic deformation}\label{elastic_defor}

Elastic deformation is a data augmentation technique used in computer vision that involves deforming an image by stretching and compressing it in random directions. This is achieved by applying small displacements to each pixel in the image. The displacements are computed using a random displacement field, which is generated by convolving a smooth noise field with a Gaussian filter. The amount of deformation applied to each pixel is controlled by a scaling factor that determines the strength of the deformation\cite{hussain2017differential, xu2020automatic, kim2022deep, zhong2021deep}. Elastic deformation can help the model become more robust to deformations in the tissue. In computational pathology, tissues may undergo various types of deformations, such as stretching or compression, due to the way they are prepared for analysis. By applying elastic deformation to the images during training, the model can learn to recognize the same tissue patterns despite these deformations.

\subsection{Grid distortion}

Grid distortion is a data augmentation technique that involves distorting the image by warping a grid overlaid on it. The grid can be a regular grid of squares or a mesh grid of triangles. The vertices of the grid are randomly displaced in both x and y directions. This creates a deformation effect on the image where the pixels are moved and stretched in various directions, simulating deformations that may occur in the tissue. It is also important to keep in mind that if distortions are applied to a very high degree, they could generate artifacts that the model learns as features\cite{lee2021principled, kim2022deep, zhong2021deep}.

\subsection{Cutout}

Cutout is a data augmentation technique that involves randomly removing a square or rectangular portion of an image and replacing it with a constant value, usually zero. This technique is similar to random erasing, but instead of replacing the erased portion with noise, it is replaced with a constant value. Cutout helps the model become more robust to occlusions in the tissue, as it simulates the presence of missing or damaged tissue in the image\cite{rebuffi2021data}.

\subsection{Gabor filtering}

Gabor filtering is a data augmentation technique that involves applying a set of Gabor filters to an image. Gabor filters are a class of linear filters that are commonly used in image processing and computer vision. They are designed to detect edges and other features in an image by analyzing the variations in the intensity of the image at different spatial frequencies and orientations. The Gabor filter is a convolution kernel that is defined by a sinusoidal wave that is modulated by a Gaussian function. The sinusoidal wave captures the spatial frequency of the image, while the Gaussian function captures the local structure and texture of the image. By applying a set of Gabor filters with different spatial frequencies and orientations to an image, the resulting filtered images can capture a wide range of features and structures in the image. In the context of data augmentation, Gabor filtering can be used to generate new images that are similar to the original image but have different features and structures highlighted\cite{barshooi2022novel}.

\subsection{Sobel filtering}

Sobel filter is a popular edge detection filter used in image processing. In the context of data augmentation, the Sobel filter can be used to generate new images with edge features. To apply the Sobel filter as a data augmentation technique, the filter is convolved with the input image. The Sobel filter is a 3x3 matrix with specific values that are multiplied with the pixel values in the image. This process calculates the gradient of the image intensity at each pixel and highlights the edges in the image. The filter is applied in both the x and y directions to capture horizontal and vertical edges. After applying the Sobel filter to an input image, the resulting image contains edge information in the form of gradient magnitude and orientation. This can be used to augment the training data by creating new images with edge features. The augmentation can be performed by adding the Sobel filtered image to the original image, multiplying it by a scaling factor, or concatenating it with the original image\cite{barshooi2022novel}.

\subsection{Random erasing}

Random erasing is a data augmentation technique that involves randomly selecting a rectangular patch in an image and replacing it with noise. This process simulates occlusions or missing information in the image, which can occur due to staining artifacts, folds in the tissue, or other factors. By introducing these types of distortions into the training data, the model can become more robust to such occlusions and learn to recognize the underlying features of the tissue that are not affected by these artifacts. Random erasing can be applied to both color and grayscale images and can be controlled by parameters such as the probability of erasing a patch, the size of the patch, and the type of noise used to replace it. Overall, the use of random erasing can improve the generalization performance of the model by allowing it to learn to recognize tissue patterns even in the presence of occlusions\cite{perez2018data}.

\section{Data mixing based}

\subsection{Mixup}

Mixup is a data augmentation technique that involves combining two or more images by taking a weighted average of their pixels to create a new synthetic image. The weight assigned to each image determines the contribution it makes to the final image. For example, if two images are mixed with weights of 0.5 each, the resulting image will have pixel values that are the average of the corresponding pixels in the two original images\cite{zhang2021carvemix, rebuffi2021data}.

\subsection{CutMix}

CutMix is a data augmentation technique that is similar to the Mixup technique, but instead of taking a weighted average of the pixels of two or more images, a portion of one image is cut and replaced with a portion of another image. This portion is selected at random from a bounding box that covers a certain percentage of the image area. The bounding box can be of any shape and size, and it does not have to be rectangular\cite{zhang2021carvemix, rebuffi2021data}.

\subsection{CarveMix}
This data augmentation technique is similar to other "mix"-based methods like Mixup and CutMix, which combine two labeled images to create new labeled samples. However, unlike these methods, CarveMix is lesion-aware, meaning that the combination is performed with attention to the lesions and a proper annotation is created for the generated image.

To create new labeled samples using CarveMix, a region of interest (ROI) is carved out from one labeled image based on the location and geometry of the lesion, and the size of the ROI is sampled from a probability distribution. The carved ROI is then inserted into the corresponding voxels of a second labeled image, and the annotation of the second image is updated accordingly. This generates new labeled images for network training while preserving the lesion information \cite{zhang2021carvemix}.

\subsection{Style transfer}

Style transfer is a data augmentation technique that involves transferring the style of one image onto another image while preserving its content. This technique is based on the concept of neural style transfer, which uses deep neural networks to extract the style and content features of images. In the context of computational pathology, style transfer can be used to generate new images that have similar texture or appearance as the original images but with different styles. For example, the style of a high-quality image can be transferred onto a low-quality image to improve its appearance and make it more suitable for analysis. Style transfer can also be used to generate synthetic slide images that mimic the appearance of real tissue samples with a specific pathology stain, which can be useful for training deep learning models on a larger and more diverse dataset\cite{shin2021style, mukherkjee2022brain}.

\subsection{CycleGAN}

CycleGAN is a data augmentation technique that uses Generative Adversarial Networks (GANs)  to translate images from one domain to another. CycleGAN is called "cycle" because it is trained using cyclic consistency loss, which means that the model is trained to translate an image from one domain to another and then back again to the original domain, while still maintaining the original content of the image. This cycle consistency ensures that the model produces high-quality and realistic synthetic images that resemble the images from the target domain. In the context of computational pathology, CycleGAN can be used to translate images from one type of tissue to another, or from one imaging modality to another. For instance, it can be used to generate synthetic images of a particular type of tissue with different staining protocols or from different scanners, which can help the machine learning algorithms to learn to recognize the tissue irrespective of the staining protocol or scanner used. This can be especially useful when working with large and complex datasets where there may be variations in image quality or staining protocols\cite{xu2019semi, liu2021ct, vasiljevic2022cyclegan}.

\subsection{Super-resolution}

Super-resolution is a data augmentation technique that involves increasing the resolution of an image, such that the resulting image has a higher pixel density and appears more detailed and sharper than the original image. This is achieved by using machine learning algorithms that are trained on pairs of low-resolution and high-resolution images to learn how to generate high-resolution images from low-resolution inputs. This can be particularly useful in cases where the dataset contains a limited number of high-resolution images, or where it is difficult or expensive to obtain high-resolution images. By generating synthetic high-resolution images, the dataset can be augmented with additional images that have more detailed information, thereby improving the performance of machine learning models that are trained on the dataset\cite{isaac2015super, kaji2019overview}.

\section{Filters and masks}

\subsection{Unsharp mask}

There are several algorithms that can be used for sharpening an image, but the most common approach is to apply a filter that accentuates the high-frequency components of the image. One popular filter for this purpose is the unsharp mask (USM) filter, which works by subtracting a blurred version of the image from the original image. This enhances the edges and details by increasing the contrast of the pixels surrounding them\cite{myint2018analysis, kucs2013detection}.

\subsection{Laplacian filter}

Another sharpening technique is the Laplacian filter, which is based on the second derivative of the image. This filter enhances the edges by detecting regions of rapid intensity change, which are likely to correspond to edges or details in the image. The resulting image, if used as data augmentation, will have the edges enhanced\cite{abdallah2015augmentation}.

\subsection{Blurring}

Blurring  is a data augmentation technique that is used to reduce the high-frequency components in an image, effectively smoothing out the image. The blurring process involves convolving the image with a kernel, which is essentially a small matrix of numbers. The kernel is typically a low-pass filter that attenuates the high-frequency components in the image while preserving the low-frequency components. The degree of blurring can be controlled by adjusting the size of the kernel, with larger kernels resulting in more severe blurring. Blurring can be useful in various applications besides data augmentation, such as denoising, image compression, and feature extraction. For example, in computational pathology, blurring can be used to remove noise or other unwanted artifacts that can interfere with the accurate detection of structures of interest. In addition, blurring can also be used to extract features from an image by highlighting the low-frequency components, such as edges or texture patterns, while suppressing the high-frequency noise\cite{sirazitdinov2019data, deoliveira2021classification}.

\subsubsection{Gaussian blur}

Gaussian blur is a type of blur that applies a Gaussian filter to the image. The Gaussian filter is a bell-shaped curve that is used to blur the image by smoothing out the pixels. This type of blur is commonly used in image processing because it is computationally efficient and produces a smooth, visually appealing effect. The amount of blur can be controlled by adjusting the size of the Gaussian filter kernel\cite{hussain2017differential, tellez2019quantifying}.

\subsubsection{Median blur}

Median blur, on the other hand, works by replacing each pixel in the image with the median value of the neighboring pixels. This type of blur is useful for removing salt-and-pepper noise from the image, which can occur when individual pixels in the image are randomly set to either the maximum or minimum value. Median blur can preserve edges in the image better than other types of blur, but it can be slower to compute\cite{akarsu2016fast, sun1994detail}.

\subsubsection{Bilateral blur}

Bilateral blur is a type of blur that preserves edges while blurring the rest of the image. It works by applying a Gaussian filter to the image, but with an additional parameter that controls how much the filter should take into account the difference in intensity between neighboring pixels. This means that the filter will blur the image, but will also preserve the edges where there are sharp changes in intensity. Bilateral blur can be slower to compute than other types of blur, but it is often used in cases where edge preservation is important\cite{bhonsle2012medical}.

\subsection{Local Binary Pattern}

Local Binary Pattern (LBP) is a texture descriptor that captures the local structure of an image. It is a simple yet powerful technique that can be used to extract texture features from an image. LBP works by comparing the gray-level intensity values of neighboring pixels in a circular region around a central pixel. The result is a binary code that represents the pattern of the intensities in the neighborhood. In order to calculate the LBP, it is necessary to perform the following calculations. First, a threshold is set based on the intensity value of the central pixel. Then, the surrounding pixels are compared to the threshold. If the intensity of the surrounding pixel is greater than or equal to the threshold, it is assigned a binary value of 1. If the intensity is less than the threshold, it is assigned a binary value of 0. This process is repeated for each surrounding pixel, resulting in a binary code. The binary codes are then combined to form a decimal value, which represents the LBP feature for that pixel. This process is repeated for all pixels in the image, resulting in an LBP map that describes the texture of the image\cite{mustaqim2022wavelet, mubarak2022local, maheshwari2022comparison}.

To apply LBP as a data augmentation technique, the LBP feature is calculated for each pixel in the image. This results in a new image where each pixel is replaced by its LBP feature value. This transformed image can then be used as an additional training sample to improve the robustness of the model to variations in texture and local structure.

\section{Division based }

\subsection{Patch-based augmentation}

Patch-based augmentation is a technique where an image is divided into smaller patches, and data augmentation techniques are applied to each patch individually. The process starts with dividing the original image into smaller patches, usually with the same size, shape, and number of pixels. Each patch is treated as a separate image and subjected to various data augmentation techniques, such as rotation, scaling, flipping, cropping, or noise addition. The choice of augmentation technique depends on the specific task and the type of data being processed. After applying the data augmentation techniques to each patch, the patches are combined to reconstruct the augmented image. This results in a more diverse training set, as each patch has undergone a different set of transformations. 
Although this technique can help with overfitting and dataset diversity it can also increase the computational complexity of the training process, as each patch must be processed individually. Moreover, the choice of patch size can also affect the performance of the model, as smaller patches can capture more local details, while larger patches can capture more global features\cite{liu2016patch, qin2021understanding}.

\subsection{Image segmentation-based augmentation}

Image segmentation-based augmentation is a data augmentation technique that involves applying data augmentation techniques to individual segments of an image based on their semantic class. In this technique, the first step is to perform image segmentation to identify different regions or objects in the image. This can be done using various techniques, such as thresholding, clustering, or deep learning-based segmentation methods. Once the different regions or objects have been identified, data augmentation techniques can be applied to each segment separately based on its semantic class\cite{olsson2021classmix}. An example of applying this technique in computational pathology is to enhance the contrast of tumoral regions in a tissue sample through data augmentation. Similarly, in cell segmentation tasks, the image can be segmented into individual cells, and data augmentation techniques can be applied to each cell separately based on its semantic class (cell or non-cell), helping the model become more robust to variations in cell shape and texture.

\section{Multi-scale and multi-view based}

\subsection{Multi-scale training}\label{multiscale_tr}

Multi-scale training is a data augmentation technique used to train deep neural networks on images of varying resolutions. In this technique, the training images are resized to multiple resolutions, and the model is trained on each of these resized images. The primary goal of multi-scale training is to make the model more robust to variations in image size. In many real-world scenarios, images can have varying resolutions, which can make it challenging for the model to generalize well. By training the model on images of different sizes, the model can learn to recognize objects and patterns at different scales and be more adaptable to variations in image size. Multi-scale training can be implemented in various ways. One approach is to randomly resize the training images to different resolutions during each training iteration. Another approach is to train the model separately on images of different resolutions and then combine the models' outputs during testing\cite{haber2018learning, poudel2021deep, peng2021multiscale}.

At first glance, one could interpret this technique as similar to zooming (See section \ref{zooming}). However, they are different. Zooming involves randomly zooming in or out of the image, which changes the scale of the objects in the image. Multi-scale training, on the other hand, involves training the model on images of different resolutions. Increasing resolution does not change the scale of the image, but it does change the level of detail and the amount of information present in the image. In other words, increasing the resolution means increasing the number of pixels in the image, which can make it possible to see more details in the image at the same scale.

\subsection{Multi-view training}

Multi-view training is a data augmentation technique used to train machine learning models on images captured from different viewpoints. The idea is to capture the same object or tissue from multiple angles, thus providing a diverse set of training examples that can help the model become more robust to variations in tissue orientation. Multi-view training can be achieved by acquiring images from different angles during data collection or by applying image transformations such as rotations and flips during the training process. By training the model on multiple views, the model can learn to recognize features and patterns from different perspectives, leading to improved generalization and robustness in real-world scenarios\cite{zhao2018multi, yuan2019automatic, carneiro2015unregistered, zhao2018towards}.

\subsection{Progressive resizing}

Progressive resizing is a technique used in deep learning for computer vision tasks, such as image classification and object detection. It involves starting the training with small images and gradually increasing their size during the training process. The idea behind progressive resizing is to first train the model on low-resolution images and then increase the image resolution as the training progresses. This allows the model to learn low-level features such as edges and corners on small images and then gradually learn higher-level features such as patterns and shapes on larger images. By starting with small images, the model learns to generalize better and is less likely to memorize specific details of the training data. As the image size increases, the model is forced to learn more complex features, which can result in better performance on larger images. Additionally, progressive resizing requires more computational resources and longer training time as the image size increases. Therefore, it is important to balance the benefits of progressive resizing with the practical limitations of the available resources\cite{bhatt2021cervical, bhatt2021covid}.

This technique can appear similar to multi-scale training (See section \ref{multiscale_tr}). However, the main difference between progressive resizing and multi-scale training is in how the different resolutions of images are used during training. In progressive resizing, the model is trained on low-resolution images first, and then gradually larger images are used during subsequent training epochs. On the contrary, in multi-scale training, the model is trained on images of different resolutions, usually by randomly selecting the resolution of the input image for each training example.

\section{Meta-learning based}

\subsection{Neural augmentation}

The proposeddata augmentation technique involves the utilization of a two-stage process for enhancing the performance of a classification neural network. In the first stage, two images extracted from the training set are passed through a neural network (Neural Style Transfer) referred to as an "augmenter," which generates a third image that may exhibit similar style or context to the original training images. This newly created image is considered as an augmented image, and it is subsequently combined with the other images from the training set and passed through a second neural network, which serves as a classifier to perform the classification task and calculate a training loss. The training loss is then backpropagated through the classifier and augmenter neural networks to update their respective parameters. Specifically, the augmenter network is optimized to generate augmented images that lead to a lower training loss, thereby enhancing the accuracy of the classifier. Through repeated cycles of this process, the augmenter network progressively generates augmented images that are increasingly aligned with the classification task, leading to improved performance of the overall system\cite{perez2017effectiveness, shorten2019survey}.

\subsection{Smart augmentation}

The smart augmentation technique shares similarities with the neural augmentation technique, as discussed earlier. However, a key difference lies in the use of an alternate convolutional neural network (CNN) instead of style transfer. The approach entails the use of two distinct networks, with the first one acting as an augmented image generator, similar to the previous technique. This generator processes two or more images and produces new ones. The second network, on the other hand, carries out the classification task by taking both the original training set and the generated images as input for training.

The training loss is then backpropagated by both the first and the second networks, ensuring that the augmented images contribute to optimizing the loss function. Additionally, the first network also employs its own loss function, which guarantees that the generated images are similar to the class from which they originate\cite{shorten2019survey}.

\subsection{Auto augmentation}

The auto-augmentation technique also endeavors to identify the most effective augmentations automatically, using reinforcement learning. In this context, reinforcement learning involves a policy that encompasses all feasible actions, or options, that the agent (augmentations generator) can take to achieve a particular objective. In the case of auto-augmentation, the policy comprises of sub-policies, where each sub-policy represents a particular type of augmentation along with its magnitude, such as a 90-degree rotation. Thus, the agent explores all the sub-policies or augmentation options to minimize the loss function of a classifier, which is its primary objective\cite{shorten2019survey}.

\section{Other techniques}

\subsection{Random patches}

Random patches augmentation is a data augmentation technique where random patches are extracted from the original image and used as new training samples. This technique can be used in different ways, such as selecting random patches from the original image with a fixed size, selecting random patches with a size proportional to the original image size, or selecting patches that cover a certain percentage of the original image. It is important to ensure that the selected patches are still representative of the original image and do not introduce biases in the training data. 
There is difference between generating random patches and random cropping. Random cropping involves randomly selecting a subregion of an image and using it as a new sample. The cropped subregion can be of any size and aspect ratio, and it can be located anywhere in the original image. The generation of random patches, on the other hand, involves randomly selecting two or more non-overlapping patches from different images to expand a dataset. Sometimes patches from different images can also be combined to generate a new image\cite{takahashi2019data}.

\section{Final remarks}

The preceding paragraphs provide an exhaustive review of the augmentation techniques used in computer vision for medical imaging. It is observed that some techniques rely entirely on manual selection of parameters by the user, whereas others incorporate a certain degree of automation, allowing for the transfer of styles and the generation of highly realistic artificial images. Furthermore, fully automatic techniques employ a linked pair of networks, with one generating the augmentations and the other assessing their suitability for the given task. This process alters the types and magnitude of augmentations in each training cycle, without requiring user intervention. These techniques enable the development of more robust models that can be applied in domains with limited or challenging data availability. It is anticipated that the list of available techniques will expand in the future, providing researchers with additional options to consider.

\bibliography{ref.bib}

\bibliographystyle{IEEEtran}

\vspace{12pt}

\end{document}